\documentclass{elsart}
\usepackage{ifpdf}
\usepackage{graphicx,amssymb,lineno}
\usepackage{bm}
\ifpdf
\usepackage[%
  pdftitle={Instructions for use of the document class
    elsart},%
  pdfauthor={Simon Pepping},%
  pdfsubject={The preprint document class elsart},%
  pdfkeywords={instructions for use, elsart, document class},%
  pdfstartview=FitH,%
  bookmarks=true,%
  bookmarksopen=true,%
  breaklinks=true,%
  colorlinks=true,%
  linkcolor=blue,anchorcolor=blue,%
  citecolor=blue,filecolor=blue,%
  menucolor=blue,pagecolor=blue,%
  urlcolor=blue]{hyperref}
\else
\usepackage[%
  breaklinks=true,%
  colorlinks=true,%
  linkcolor=blue,anchorcolor=blue,%
  citecolor=blue,filecolor=blue,%
  menucolor=blue,pagecolor=blue,%
  urlcolor=blue]{hyperref}
\fi

\makeatletter
\def\elsartstyle{%
    \def\normalsize{\@setfontsize\normalsize\@xiipt{14.5}}
    \def\small{\@setfontsize\small\@xipt{13.6}}
    \let\footnotesize=\small
    \def\large{\@setfontsize\large\@xivpt{18}}
    \def\Large{\@setfontsize\Large\@xviipt{22}}
    \skip\@mpfootins = 18\p@ \@plus 2\p@
    \normalsize
}
\@ifundefined{square}{}{\let\Box\square}
\makeatother

\begin{document}

\begin{frontmatter}
\title{Extreme vorticity growth in Navier-Stokes turbulence}

\author{J\"org Schumacher $^1$, Bruno Eckhardt$^2$, and Charles R. Doering$^3$}
\address{$^1$ Institut f\"ur Thermo- und Fluiddynamik, Technische Universit\"at Ilmenau, Postfach 100565, D-98684 Ilmenau, Germany,\\
$^2$ Fachbereich Physik, Philipps-Universit\"at Marburg, D-35032 Marburg, Germany,\\
$^3$ Departments of Mathematics and Physics, and Center for the Study of Complex Systems,
          University of Michigan Ann Arbor, Ann Arbor, MI 48109-1043, USA}

\begin{abstract}
According to statistical turbulence theory, the ensemble averaged squared vorticity $\rho_E$ 
is expected to grow not faster than $\mbox{d}\rho_E/\mbox{d}t\sim \rho_E^{3/2}$. Solving a variational 
problem for maximal bulk enstrophy ($E$) growth,  velocity fields were found for which the growth rate is as 
large as $\mbox{d}E/\mbox{d}t\sim E^3$. Using numerical simulations with well resolved small scales and a 
quasi-Lagrangian advection to track fluid subvolumes with rapidly growing vorticity, we study spatially resolved 
statistics of vorticity growth. We find that the volume ensemble averaged growth bound is satisfied locally to 
a remarkable degree of accuracy. Elements with $\mbox{d}E/\mbox{d}t \sim E^3$ can also be identified, 
but their growth tends to be replaced by the ensemble-averaged law when the intensities become too large. 
\end{abstract}

\end{frontmatter}

\section{Introduction}
\label{intro}
Fluid turbulence may be characterized as a tangle of intermittent vortices embedded in regions of straining 
motion \cite{Orszag1991,Ishihara2007}. Theoretical and computational studies of the evolution of isolated 
intense vortices \cite{Burgers1948,Batchelor1964}, pairs of vortices 
\cite{Lundgren1982,Kerr1985,Pullin1998,Horiuti2008}, highly-symmetric vortex tangles or ensembles of randomly distributed 
vortices \cite{Boratav1994,Kambe2000} 
have contributed considerably to our understanding of the statistical properties of homogeneous isotropic 
Navier-Stokes turbulence, but numerical computation of high-vorticity events in a turbulent flow remains elusive because 
of the high demands on spatial and temporal resolution. The properties of such events, their frequency and 
maximal intensities, are important for small scale mixing, the efficiency of combustion processes, and for 
modeling turbulence. The situation for the inviscid Euler equation is not much different. 
Recently, Bustamente
and Kerr \cite{Bustamente2008} discussed in detail the sensitivity of vorticity growth on grid resolution and de-aliasing techniques for interacting highly-symmetric anti-parallel vortices. They came to different conclusions from Hou and 
co-workers \cite{Hou2006} who observed a depletion of the vortex stretching in their Fourier smoothing method. The question of finite-time blow-up in solutions of the Euler equations remains an open area of investigation.   

For viscous fluids there is a bound on the relation between the growth rate of the volume integrated 
squared vorticity $E$, namely $\mbox{d}E/\mbox{d}t \le a E^3$ with known prefactor $a$, and there is also a 
calculation of optimal fields for which $\mbox{d}E/\mbox{d}t \sim E^3$. On the other hand, the  
ensemble averaged squared vorticity $ \rho_E$ is not expected to grow faster than 
$\mbox{d}\rho_E/\mbox{d}t\sim \rho_E^{3/2}$. In an effort to understand the relation between these two results and how they translate to turbulence, we conducted high-resolution direct numerical 
simulations with a special focus on the dynamics on small scales and the evolution of strong vorticity 
amplification elements. This is the objective of the present work.

The strong temporal variations near ``almost singular'' events 
in turbulence can only be resolved with a sufficiently small time step, which together with a high spatial 
resolution requires the storage of huge data files. We use a cubic box with periodic boundaries in all 
directions and solve the Navier-Stokes equations numerically,
\begin{eqnarray}
\frac{\partial {\bm u}}{\partial t} + ({\bm u\cdot\bm\nabla}){\bm u} &=& - {\bm\nabla}p +
\nu \Delta {\bm u}+ {\bm f}\,,\\
\bm\nabla\cdot\bm u&=&0\,,
\label{nse}
\end{eqnarray}
where $\bm f$ is a large-scaling forcing.   We apply the pseudospectral method with 2/3 de-aliasing 
and obtain a homogeneous isotropic and statistically stationary three-dimensional flow at a Taylor microscale Reynolds 
number of $R_{\lambda}=107$ \cite{Schumacher2007}. The grid size is $2048^3$, so that when 
expressed in terms of the Kolmogorov length $\eta_K=\nu^{3/4}/ \langle\epsilon\rangle^{1/4}$ (with 
the mean energy dissipation rate $\langle\epsilon\rangle$) the sides of the box are $683 \eta_K$ 
long and there are 3 grid points per $\eta_K$. Since the crossover from viscous to inertial 
range occurs near a scale of $8\eta_K$, we can resolve singular events into 
the viscous range. The time step $\Delta t=0.003\tau_{\eta}$, where $\tau_{\eta}=\sqrt{\nu/\langle
\epsilon\rangle}$ is the Kolmogorov time, is well within the limits discussed by Donzis and 
Sreenivasan \cite{Donzis2008}. The events we would like to study are followed for a time interval
of about 55 $\tau_{\eta}$ units or 1830 output steps. This would add up to about $5\times 10^{13}$ 
velocity field values that have to be stored. In order to avoid this, we turn to the so-called 
quasi-Lagrangian method \cite{Belinicher1987} which eliminates the large scale sweeping motion 
superimposed on the localized singular events we want to study. Specifically, we follow 100 Cartesian 
cubes $V_L$ simultaneously through different regions of the evolving Navier-Stokes flow. The motion 
of the subvolumes $V_L$ is fixed by the advection of a Lagrangian tracer in their center, and their sides are kept 
aligned with the outer coordinates. The boxes have a side length of $L=17 \eta_K$, corresponding
to $51^3$ grid points. This reduces the number of velocity field values that have to be stored to 
$1.4\times 10^{-3}$ of the original estimate.

\section{Analytic predictions on the growth rate of enstrophy}
The vorticity is the curl of the velocity field, 
${\bm\omega}={\bm\nabla\times \bm u}$, and the enstrophy is the volume integral of its intensity,
\begin{equation}
E(t)=\int_V {\bm\omega}^2 \mbox{d}V\,.
\label{enstrophy}
\end{equation}
It follows from the Navier-Stokes equation that in any incompressible viscous Newtonain fluid 
the growth rate of the enstrophy, $\mbox{d}E/\mbox{d}t$, obeys (\cite{Lu2006,LuDoering2008}),
\begin{equation}
\frac{\mbox{d}E(t)}{\mbox{d}t}=2\int_V ({\bm\omega}\cdot{\bm\nabla}{\bm u})\cdot{\bm \omega}
\mbox{d}V-2\nu \int_V ({\bm\nabla}{\bm \omega})^2 \mbox{d}V\,.
\label{ebalance}
\end{equation}
From this it can be shown that $E(t)$ cannot grow faster than  
\begin{equation}
\frac{\mbox{d}E(t)}{\mbox{d}t}\le \frac{27 c^3}{16 \nu^3}E(t)^3\,,
\label{enstrophy1}
\end{equation}
with $\nu$ the kinematic viscosity of the fluid and $c=\sqrt{2/\pi}$ (for details see \cite{Lu2006,LuDoering2008}). 
This holds for the volume integrated quantity and does not make any assumptions on the flow. Incompressible
flow fields that maximize the enstrophy production, and thus the growth rate of enstrophy, were recently found by
solving an optimization problem \cite{Lu2006,LuDoering2008}. At high Reynolds number the maximum enstrophy growth rate 
(\ref{enstrophy1}) is realized by colliding, axially symmetric vortex rings.  At lower Reynolds numbers, maximum enstrophy
generation is realized by interacting Burgers vortices with $\mbox{d}E/\mbox{d}t\sim E^{7/4}$. 

There is a second analytical result that pertains to the growth rate of the ensemble averaged squared vorticity 
(or enstrophy density) 
\begin{equation}
\rho_E=\langle{\bm\omega^2}\rangle
\end{equation}
for the particular case of homogeneous and 
isotropic (box) turbulence. A direct consequence of the von K\'{a}rm\'{a}n-Howarth (KH) equation \cite{Karman1938} for 
the velocity correlations, when the volume average $\langle \cdot \rangle$ in (\ref{KH2}) agrees 
with the ensemble average that appears in the KH equation, is derived  in \cite{Rotta1972,Davidson2004} and states that
\begin{equation}
\frac{\mbox{d}}{\mbox{d} t} \rho_E =
-\frac{7 S}{3\sqrt{15}} \rho_E^{3/2} -70\nu \langle(\partial^2_x u_x)^2\rangle
\label{KH2}
\end{equation}
where $S$ is the skewness of the longitudinal velocity derivative, and $u_x$ is the $x$-component of the turbulent velocity field. It is an empirical fact that $S<0$. 
It has been observed that the skewness $S$ is basically constant for Taylor microscale Reynolds numbers $R_{\lambda} \lesssim 200$ and it grows weakly as $|S|\sim R_{\lambda}^{0.11}$  for $R_{\lambda}>200$ \cite{Sreenivasan1997,Ishihara2009}. Thus we will assume for purposes of discussion and data analysis that $\mbox{d}\rho_E/\mbox{d}t\sim \rho_E^{3/2}$ holds approximately.  This exponent is much smaller than the one in the upper bound (\ref{enstrophy1}).

In order to make the relation between the two relations more explicit, we rewrite the first one assuming that 
$E=L^3 \rho_E=L^3 \langle{\bm\omega}^2\rangle$ for a box of length $L$ and 
$\langle\epsilon\rangle=\nu\rho_E$ to bring in the Kolmogorov length. 
Then the bound suggests that
\begin{equation}
\frac{\mbox{d}}{\mbox{d}t}\rho_E\le \frac{27 c^3}{16}
\left(\frac{L}{\eta_K}\right)^6
\rho_E^{3/2}\,.
\label{enstrophy1a}
\end{equation}
The key difference then is a ratio of lengths: if the sidelength $L$ of the 
volume $V$ is of the order of the Kolmogorov scale, both results imply the 
same exponent, despite the different boundary conditions and derivations.
But if the volume is larger than $\eta_K$, a stronger variation is possible: 
in a situation where the vorticity content of the volume is below the mean 
that enters the definition of the Kolmogorov length, the local dissipation length is larger than the statistical average, and a built up of entstrophy will reduce the local 
value, thereby increasing the contribution from the factor $(L/\eta_K)$.

\section{Results}
\subsection{Local quasi-Lagrangian analysis}
\begin{figure}
\begin{center}
\includegraphics[angle=0,scale=0.4,draft=false]{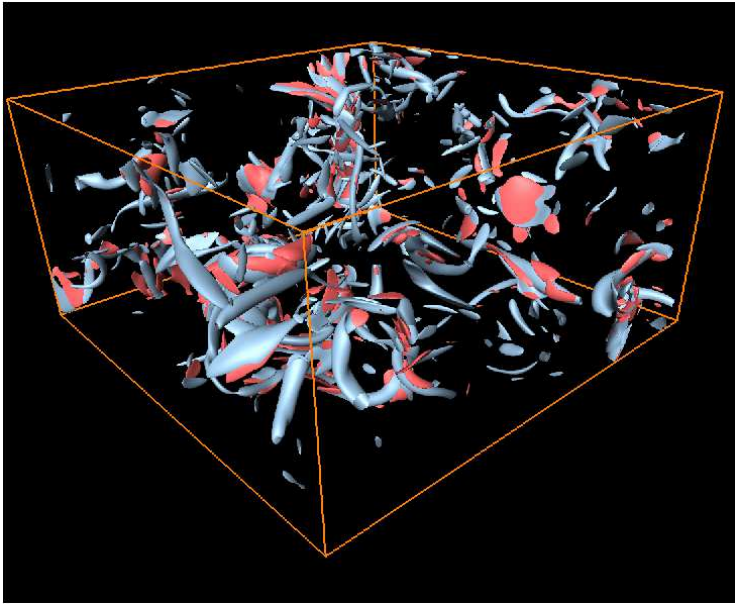}
\caption{(Color online) Vortex tubes and dissipation sheets in homogeneous isotropic turbulence.
Isosurfaces of the vorticity magnitude square (or local enstrophy) $\Omega=\omega^2$ (cyan) 
and the energy dissipation rate $\epsilon=2\nu S_{ij}S_{ij}$ (red) with $S_{ij}=(\partial 
u_i/\partial x_j+\partial u_j/\partial x_i)/2$ the rate of strain tensor. Both surfaces 
are shown at the level of ten times their mean. The displayed volume is 1/16 of the full 
simulation box.} 
\label{fig1}
\end{center}
\end{figure}

Configurations as highly symmetric as the colliding vortex rings that realize the maximum
instantaneous enstrophy generation cannot be generically expected in a 
turbulent flow.  In particular, the high-amplitude vorticity events in turbulence arise 
preferentially in the form of tubes \cite{Orszag1991,Ishihara2007,Jimenez1998} which 
are rapidly stretched and deformed by background and self-induced straining motions 
\cite{Hamlington2008,Hamlington2008a}. Strong strain (or shear) results in by high-amplitude energy 
dissipation rates. The spatial distribution of high-amplitude events of the local enstrophy 
and energy dissipation rate, illustrated in Figure \ref{fig1}, underlines this behavior. 
Isosurfaces of the {\em local} enstrophy, $\Omega({\bm x},t)={\bm \omega}^2({\bm x},t)$, 
(cyan) at ten times the mean value show primarily elongated structures. The isosurfaces 
of energy dissipation rate, $\epsilon({\bm x},t)=(\nu/2) (\partial u_i/\partial x_j+\partial u_j/
\partial x_i)^2$, (red) also at ten times the mean value, reveal sheet-like structures between 
the high-vorticity events. This illustrates that while the ensemble averaged values of energy 
dissipation and vorticity are related by $\langle\epsilon\rangle=\nu\rho_E$ this does 
not apply to their instantaneous and local values. It does show, however, that extreme 
events occur at neighboring locations.
\begin{figure}[t]
\begin{center}
\includegraphics[angle=0,scale=0.55,draft=false]{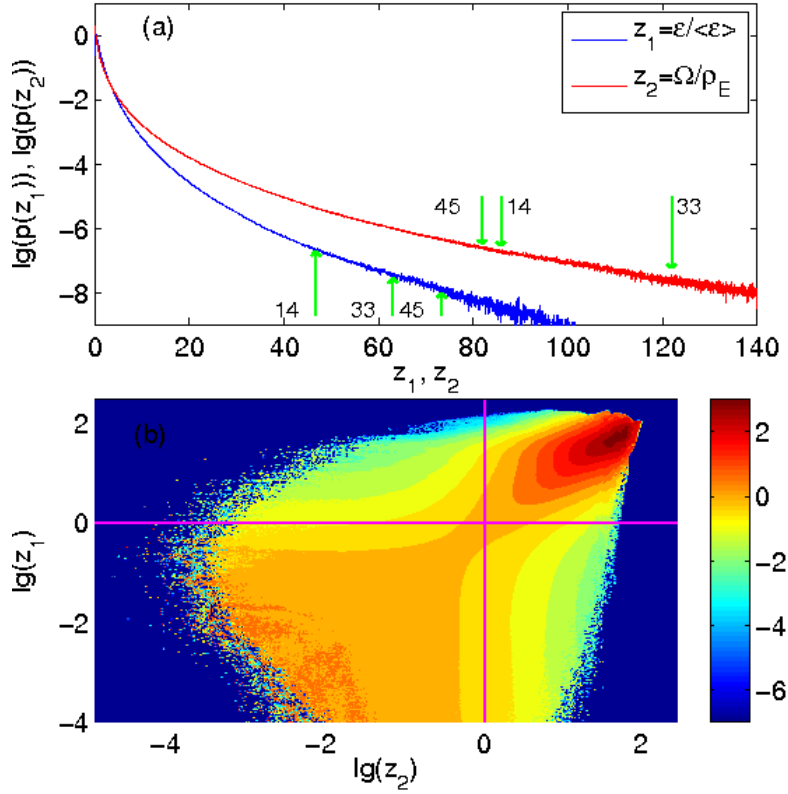}
\caption{(Color online) Statistics of local enstrophy and energy dissipation rate. (a) Probability 
density functions of local enstrophy, $\Omega$, and energy dissipation rate, $\epsilon$, given 
in units of their means, respectively. The vertical arrows mark the global maxima of $\Omega$ 
and $\epsilon$, respectively, in case of three Lagrangian tracers, no. 14, 33 and 45. 
(b) Joint probability density function of local enstrophy and energy dissipation rate. 
The distribution is normalized by both single quantity distributions, $p(z_1,z_2)/[p(z_1)p(z_2)]$, 
in order to highlight the statistical correlations between $z_1=\epsilon/\langle\epsilon\rangle$ 
and $z_2=\Omega/\rho_E$. Color coding is in decadic logarithm.} 
\label{fig2}
\end{center}
\end{figure}
\begin{figure}[t]
\begin{center}
\includegraphics[angle=0,scale=0.6,draft=false]{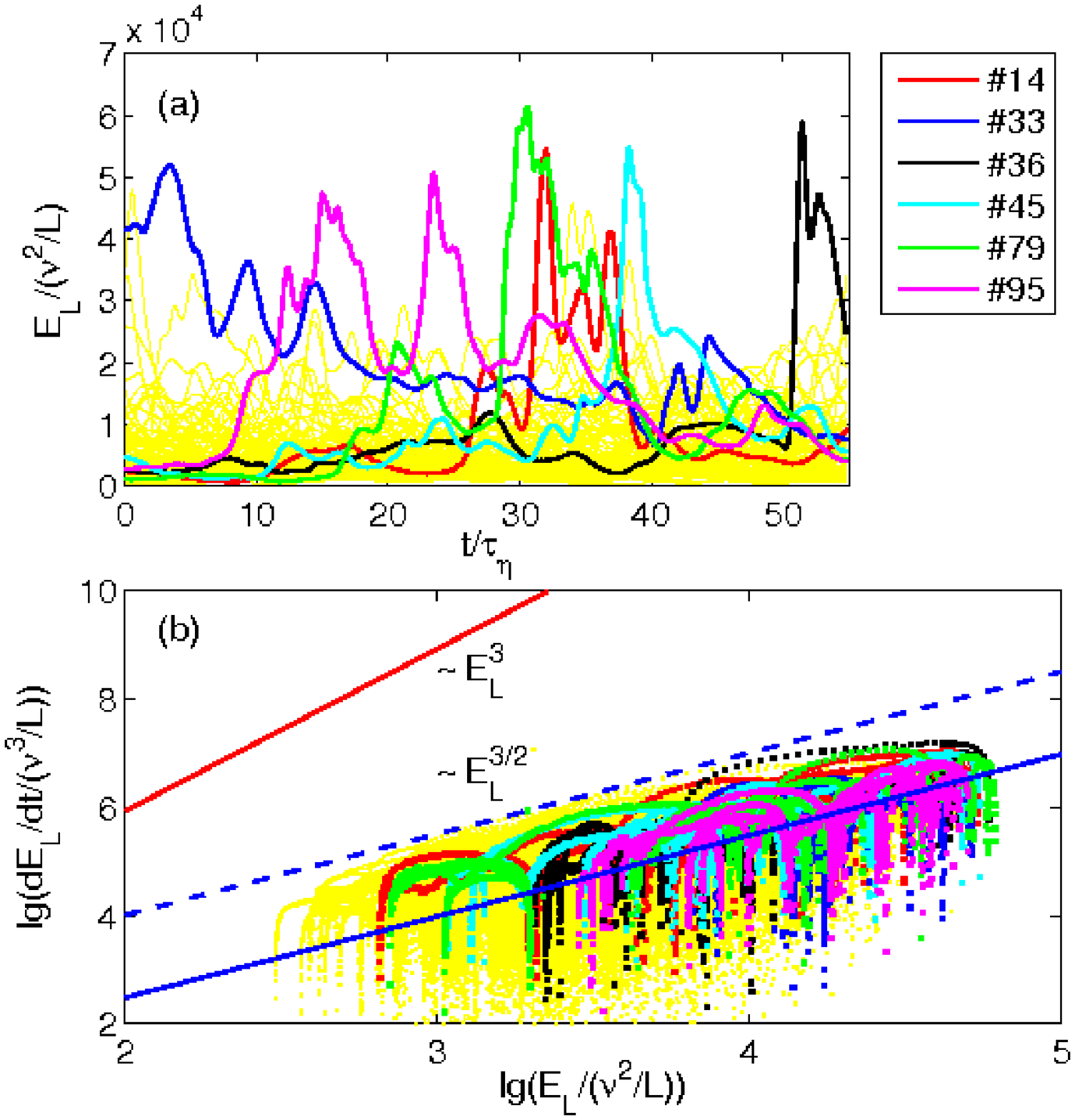}
\caption{(Color online) Quasi-Lagrangian analysis of enstrophy. (a) Time series of 
$E_L(t)$ for all 100 subvolumes $V_L$ are plotted. The time traces that reach the largest 
local maxima for $E_L$ are colored differently and their labels are indicated 
in the legend. Enstrophy is given in units of $\nu^2/L$ with $L=16\eta_K$. (b) 
Enstrophy growth rate versus enstrophy. The enstrophy growth rate, $\mbox{d}E_L/
\mbox{d}t$, is given in units of $\nu^3/L^3$. The {\it a priori} upper bound 
$\mbox{d}E_L/\mbox{d}t=27\sqrt{2}/(8\nu^3\sqrt{\pi^3}) E_L^3$ is indicated 
as a red line. The growth that follows from the von K\'{a}rman-Howarth equation 
\cite{Karman1938}, $\mbox{d}E_L/\mbox{d}t \approx -7S/3\sqrt{15} E_L^{3/2}$ 
with a derivative skewness of $S=-0.5$, is indicated as a solid blue line. 
The dashed blue line has the same slope and serves as a guide to the eye.
Color coding is as in panel (a).} 
\label{fig3}
\end{center}
\end{figure}

To capture this quantitively, we study the probability distribution functions (pdf's) of local energy
dissipation and vorticity in Figure \ref{fig2}. The pdf's of  $\epsilon({\bm x},t)$ and $\Omega({\bm x},t)$ 
show stretched exponential tails indicative of strong small-scale intermittency (see Figure 2(a)). 
The tail is more extended for $\Omega$ than for $\epsilon$, in agreement with the observations in 
\cite{Zeff2003}. Fat tails imply that large amplitude events are significantly more probable than for 
a Gaussian distributed signal with the same second moment. In Figure 2(a) we also mark the global 
maxima in $\epsilon$ and $\Omega$ identified within the advected volumes $V_L$ for three particular 
Lagrangian trajectories. The locations far out in the tails document that our quasi-Lagrangian tracking 
is able to detect high-amplitude events and that extreme events in both quantities are spatially 
correlated and located within our advected volume. The local correlation between high-amplitude 
local enstrophy and energy dissipation events is further supported by the joint pdf in Fig. \ref{fig2}(b),  
where $p(\epsilon/\langle\epsilon\rangle,\Omega/\rho_E)/ [p(\epsilon/\langle\epsilon
\rangle) p(\Omega/\rho_E)]$ is shown. The maximum values appear in the upper right 
of the support where the largest amplitudes for both are present. High-amplitude fluctuations in 
energy dissipation and local enstrophy density are thus strongly statistically correlated and found 
very close together, in both, space and time. 

We now turn to the study of the time evolution of extreme events within our subvolumes.
Figure \ref{fig3}(a) shows time traces of the local enstrophy 
\begin{equation}
E_L(t)=\int_{V_L}{\bm \omega}^2 \mbox{d}V\,.
\end{equation}
Since we are interested in the relation between large values of $\mbox{d}E_L/\mbox{d}t$ 
with $E_L$, we show in Fig.~\ref{fig3}b the same data as a scatter plot in the plane spanned by $E_L$ and $\mbox{d}E_L/\mbox{d}t$ on a double logarithmic scale. 
The collection of the individual growth histories in the subvolumes (which can vary strongly from one to another) is bounded from above by the scaling $\mbox{d}E_L/\mbox{d}t \lesssim E_L^{3/2}$, indicated there by the dashed line.
This shows that the local 
growth rate and enstrophy are related very much as are the volume averages, Eq. (\ref{ebalance}), 
so that the effects of sweeping are averaged out. It suggest that the estimate
\begin{equation}
\frac{\mbox{d}E_L}{\mbox{d}t}\sim \int_{V_L} ({\bm\omega}\cdot{\bm\nabla}{\bm u})\cdot{\bm \omega}
\mbox{d}V\sim \langle{\bm\omega}^2\rangle^{3/2} V_L \sim \sqrt{E_L^3/V_L}\,.
\label{enstrophy4}
\end{equation}
also holds locally. We have also verified this scaling in a volume $V_L/8$, i.e. in 
a cube with half the sidelength. The striking observation is that the envelope of the 
local enstrophy growth follows the scaling of (\ref{KH2}). 
However, we wish to stress that (generically) there are nonvanishing enstrophy fluxes, $\int_{\partial V_L} \omega^2\,\bm u\cdot \mbox{d} \bm A \ne 0$, across the bounds of $V_L$. 
Therefore (\ref{enstrophy4}) is only a heuristic estimate.     

\subsection{Local analysis of extreme events}
\begin{figure}[t]
\begin{center}
\centerline{\includegraphics[angle=0,scale=0.5,draft=false]{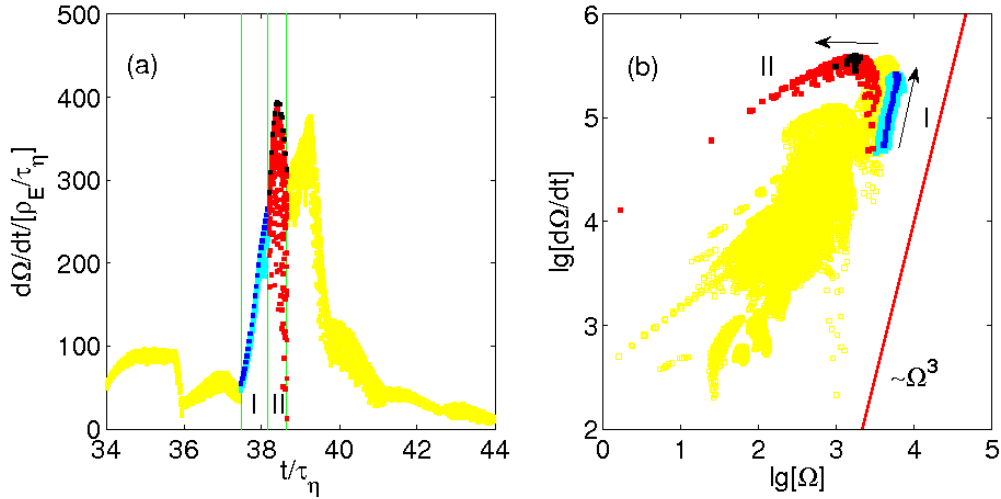}}
\caption{(Color online) Local analysis of very rapid enstrophy growth events. (a) Time evolution of the 
maximum of $\mbox{d}\Omega/\mbox{d}t$ in $V_L$ for tracer No. 45. Growth rates at the grid point of the 
maximum (blue for I and black for II) and the 27 neighbouring points (cyan for I and 
red for II) are shown. The curve is piecewise continuous since one local maximum in $V_L$ takes over a former at a 
different grid point. (b) Replot of the data from (a) in the $\mbox{d}\Omega/\mbox{d}t$--$\Omega$ 
plane. The arrows in panel (b) indicate the time evolution.} 
\label{fig4}
\end{center}
\end{figure}

The previous analysis shows that when the 
average volume has a diameter of about the Kolmogorov length or above, the
bulk estimate for the enstrophy growth is recovered. Stronger growths, therefore, 
may only occur on smaller scales. To probe for this, we turn to the study of the 
time evolution of local extrema within the cell.  We select 
the grid point ${\bm x}^*$ with the fastest local growth rate within the subvolume, 
$\mbox{d}\Omega/\mbox{d}t|_{max}=\max_{{\bm x}\in V_L(t)}[\mbox{d}\Omega/\mbox{d}t]$. 
As an indication of the numerical uncertainty, 
we also show the 26 growth rates for points on a $3^3$ cube surrounding ${\bm x}^*$. 
Figures \ref{fig4}(a) and (b) demonstrate that the 
maximum position always gives the outer envelope of the curve. The curve is continuous by construction, 
but its slope is discontinuous when the position ${\bm x}^*$ of the point of maximal growth rate 
jumps discontinuously within the cell. Panel (a) shows also that long periods of low 
variability and small growth rates are interrupted by short violent outbursts of the local enstrophy 
growth rate. The two subsequent intervals I and II for tracer No. 45 mark exactly such a rapid
growth event.

\begin{figure}[t]
\includegraphics[angle=0,scale=0.8,draft=false]{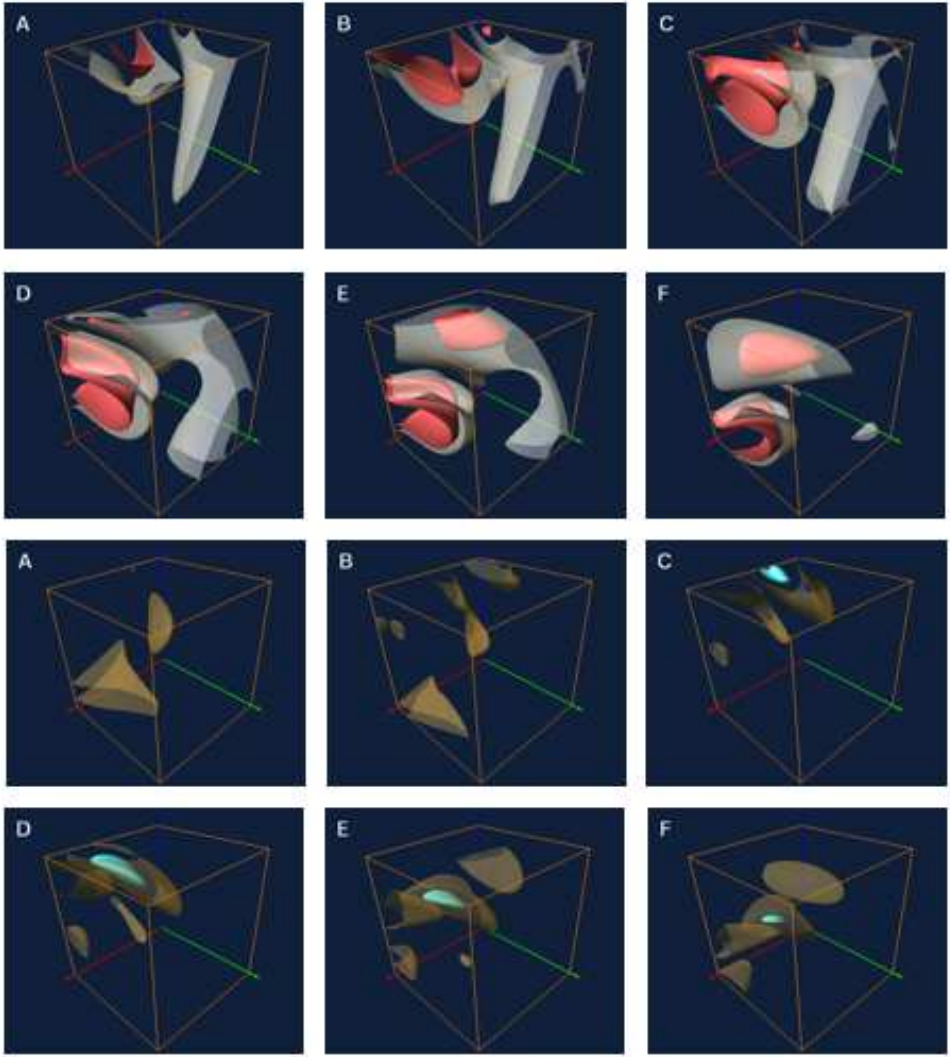}
\caption{Structures in an extreme enstrophy growth event. Upper six panels: Isosurfaces of $\Omega$ 
(red: 20 $\langle\Omega\rangle$, gray: 10 $\langle\Omega\rangle$). Lower six panels: Isosurfaces 
plots of $\epsilon$ (yellow: 2 $\langle\epsilon\rangle$, cyan: 20 $\langle\epsilon\rangle$).
Panels (A) for $t/\tau_{\eta}=$ 37.44, (B) for $t/\tau_{\eta}=37.8$, (C) for $t/\tau_{\eta}=38.16$ 
correspond with time interval I in Fig. \ref{fig4}; (D) for $t/\tau_{\eta}=38.52$, (E) for $t/\tau_{\eta}=38.88$ and
and (F) for $t/\tau_{\eta}= 39.24$ with time interval II.} 
\label{fig5}
\end{figure}

Figure \ref{fig4}(b) shows the same data as (a) in the plane spanned by the local 
enstrophy and its growth rate,  $\mbox{d}\Omega/\mbox{d}t$ and 
$\Omega$.  The first part follows a scaling that resembles the global enstrophy growth 
bound, $\mbox{d}\Omega/\mbox{d}t\sim\Omega^3$. It is connected with a rapid stretching by two 
approaching and interacting tubular vortex segments and thereby is similar to the vortex ring
collision seen for the optimal growth events. However, when dissipation kicks in 
and the high-amplitude dissipation region between the tubes forms, the growth is weaker 
as can be seen Fig. \ref{fig5} where we plot a sequence of isosurface plots of $\Omega(\bm x,t)$ and
$\epsilon(\bm x,t)$.\footnote{Although we tracked 100 subvolumes simultaneuously within the evolving turbulent flow, this specific
event was the only one observed that displayed such a rapid growth in connection with a large
local enstrophy amplitude.}
We might interpret it as a precursor of an ``attempted'' finite-time 
blowup. Further unbounded growth is prohibited by the imperfect collision of the tube segments,
since it is highly unlikely that fully developed turbulence
 generates highly symmetric flow configurations 
that pin the local vorticity maximum fixed in space \cite{Boratav1994}.

\section{Conclusions.} Our results show that the local enstrophy growth $\mbox{d}E_L/\mbox{d}t\sim E_L^{3/2}$ 
expected  from the volume and ensemble averaged Navier-Stokes equations can also be observed 
locally in volumes of a diameter equal to 17 Kolmogorov lengths that are advected by the flow.
The typical enstrophy growth time scales in these local volumes are of the order of the Kolmogorov time 
scale $\tau_{\eta}$.
They show that there is negligible influence from the sweeping motion,  so that enstrophy growth 
is dominated by local events. Within the boxes larger exponents can be observed when the total 
vorticity inside the volume is  weak,  since then the instantaneous, local vorticity gives
a larger (local) Kolmogorov length, and the increase in enstrophy and corresponding decrease in the (local)
Kolmogorov length contributes to an increasing ratio $(L/\eta_K)$.
During such events the local vorticity can be amplified with a rate $\mbox{d}E/\mbox{d}t\sim E^3$, but this growth is cut off after some $\tau_{\eta}$.
A further growth of enstrophy is then limited by the shear layer and thus the enhanced dissipation that is formed between the interacting vortex filaments. 
This conclusion is supported both by displaying the evolution of vortex filaments (as seen in Fig. 5) and by monitoring the energy dissipation rate and enstrophy accumulated in the comoving volumes (not shown).
Moreover, they seem to be rare and overwhelmed  by the $E^{3/2}$ behavior. Nevertheless, already for the  present moderate Reynolds number we found such rapid growth events and expect their more frequent appearance for larger ones. 

{\em Acknowledgements.} We thank Peter A. Davidson and Robert M. Kerr for stimulating discussions and helpful suggestions.
Supercomputing resources were provided within the DEISA consortium at the J\"ulich Supercomputing Centre (Germany). 
This work was also supported by the Heisenberg Program of the Deutsche Forschungsgemeinschaft under Grant SCHU1410/5-1 (JS), the German Academic Exchange Service (JS), the Alexander von Humboldt Stiftung (CRD), and US National Science Foundation Awards PHY-0555324 and PHY-0855335 (CRD).

\end{document}